\documentclass[aps,preprint,preprintnumbers,amsmath,amssymb,nofootinbib]{revtex4}
\usepackage{graphicx}
\usepackage{epsfig}
\begin{document}
\title{Perihelion advances for the orbits of Mercury, Earth and Pluto from Extended Theory of General Relativity (ETGR).}
\author{$^{1}$ Luis Santiago Ridao, $^{1}$ Rodrigo Avalos, $^{1}$ Mart\'{\i}n Daniel De Cicco and $^{1,2}$ Mauricio Bellini
\footnote{E-mail address: mbellini@mdp.edu.ar} }
\address{$^1$ Departamento de F\'isica, Facultad de Ciencias Exactas y
Naturales, Universidad Nacional de Mar del Plata, Funes 3350, C.P.
7600, Mar del Plata, Argentina.\\
$^2$ Instituto de Investigaciones F\'{\i}sicas de Mar del Plata (IFIMAR), \\
Consejo Nacional de Investigaciones Cient\'ificas y T\'ecnicas
(CONICET), Argentina.}

\begin{abstract}
We explore the geodesic movement on an effective four-dimensional hypersurface
that is embedded in a five-dimensional Ricci-flat manifold described by a
canonical metric, in order to applying to planetary orbits in our
solar system. Some important solutions are given, which provide
the standard solutions of general relativity without any extra
force component. We study the perihelion advances of Mercury, the
Earth and Pluto using the extended theory of general relativity
(ETGR). Our results are in very good agreement with observations
and show how the foliation is determinant to the value of the
perihelion's advances. Possible applications are not limited
to these kinds of orbits.
\end{abstract}
\maketitle

\section{Introduction, basic equations, and motivation}

The advance of the perihelion in the orbit of Mercury is a
relativistic effect\cite{Einstein}. Together with the observation
of the deflection of light by Dyson, Eddington and Davidson in
1919\cite{Dys}, this result was crucial in the final breakthrough of general
relativity. Mercury is the innermost of the four terrestrial
planets in the Solar system, moving with high velocity in the
gravitational field produced by the Sun. Because of this, Mercury
offers unique possibilities for testing general relativity and
exploring the limits of alternative theories of gravitation with
enough accuracy to be of interest\cite{Kraniotis}. A
compact calculation of the perihelion precession of Mercury in
general relativity taking into account a nonzero cosmological
Constant $\Lambda$, was considered some years ago\cite{Kra}. The same problem
was examined from five-dimensional physics, but with zero cosmological constant\cite{a}.

Lately, extensions or modifications to the standard four-dimensional
theory of general relativity have a great and increasing impact in top original researchin gravitation and cosmology.  The spectrum of these proposals
includes: theories with compact and noncompact extra dimensions\cite{Ba}, scalar-tensor theories, gravity from
non-Riemannian geometries; and $f(R)$\cite{Baibosunov}, $f(R,G)$ and $f(T)$ theories (e. g. ref them\cite{Talmadge}).

In 2009 an extended version of general relativity\cite{MB}
was introduced from a 5D Ricci-flat
space-time, where the extra space-like coordinate is noncompact.
After making a static foliation on the extra coordinate, we obtained
an effective 4D Schwarzschild-de Sitter space-time in which matter is considered with
an equation of state $\omega = p_m/\rho_m= -1$ 4D vacuum state,
such that the pressure on the effective 4D manifold is $P=-3 c^4/(
8\pi G \psi^2_0)$ and $\psi_0=c/H_0$ is the Hubble radius. The
resulting effective 4D metric is static, exterior and describes
spherically symmetric matter (ordinary matter, dark matter and
dark energy) on scales $r_0 < r_{Sch} < c/H_0$ for black holes or
$r_{Sch} < r < c/H_0$ for ordinary stars with radius $r_0$. The
radius $r_{ga}$ is very important because it delimitates distances
for which dark energy and ordinary matter (dark matter and
ordinary matter) are dominant: $r> r_{ga}$ ($r < r_{ga}$). We have 
suggested that ordinary matter, dark matter and dark
energy can be considered matter subject to a generalized
gravitational field which is attractive on scales $r< r_{ga}$ and
repulsive on scales $r > r_{ga}$.

In this work we shall study the effective 4D orbits of some
planets (or pseudo-planets in the case of Pluto) of our solar
system. In particular we are interested in the calculation of the
perihelion advances of Mercury, Earth and Pluto. In Sect. 2 we review the formalism to
calculate the orbits of massive text particles from the extended theory of general relativity (ETGR).

\section{ETGR}

In a previous work\cite{MB} a 5D extension of general relativity was
considered such that the
effective 4D gravitational dynamics had a vacuum-dominated,
$\omega =-1$, equation of state. In this section we shall examine
some formal aspects of this theory.

\subsection{5D massive test particles dynamics}

We consider the extended Schwarzschild-de Sitter 5D
Ricci-flat metric $g_{ab}$\cite{MB}
\begin{equation}\label{a1}
dS^{2}=\left(\frac{\psi}{\psi_0} \right)^{2}\left[ c^{2}f(r)dt^{2}
- \frac{dr^2}{f(r)}-r^{2}\left(d\theta ^{2}+sin^{2}(\theta)
\,d\phi ^{2}\right)\right]-d\psi^{2},
\end{equation}
where
$f(r)=1-(2G\zeta\psi_{0}/(rc^2))[1+c^{2}r^{3}/(2G\zeta\psi_{0}^{3})]$
is a dimensionless function. Here, $\psi$ is the noncompact extra
dimension. The space-like coordinates $\psi$ and $r$ have length
units, $\theta$ and $\phi$ are angular coordinates and $t$ is a
time-like coordinate. We denote $c$ the speed of light. We
shall consider that $\psi_0$ is an arbitrary constant with length
units and the constant parameter $\zeta$ has units of
$(mass)(length)^{-1}$.

For a massive free test particle outside of a spherically
symmetric compact object, the 5D Lagrangian is
\begin{equation}\label{b1}
^{(5)}L=\frac{1}{2}g_{ab}U^{a}U^{b}=\frac{1}{2}\left(\frac{\psi}{\psi_0}\right)^{2}\left[c^{2}f(r)
\left(U^t\right)^2-\frac{\left(U^r\right)^2}{f(r)}-r^{2}\left(U^{\theta}\right)^2-r^{2}sin^{2}
\theta\left(U^{\phi}\right)^2\right] -
\frac{1}{2}\left(U^{\psi}\right)^2.
\end{equation}
We shall take $\theta=\pi/2$. Because $t$ and $\phi$ are cyclic
coordinates, their associated constants of motion $p_{t}$
and $p_{\phi}$, are constants of motion. Using the five-velocity
condition $g_{ab} U^a U^b=1$, we obtain the equation of energy
for a test particle that moves on space-time (\ref{a1})
\begin{equation}\label{b6}
\frac{1}{2}\left(
U^{r}\right)^{2}+\frac{1}{2}\left(\frac{\psi_0}{\psi}\right)^{2}\left(
U^{\psi}\right)^{2} + V_{eff}(r) = E.
\end{equation}
If we identify the energy, $E$, as
\begin{equation}\label{b8}
E=\frac{1}{2}\left(\frac{\psi_0}{\psi}\right)^{4}(p_{t}^{2}c^{-2}+p_{\phi}^{2}\psi_{0}^{-2})-
\frac{1}{2}\left(\frac{\psi_0}{\psi}\right)^{2},
\end{equation}
the effective 5D potential, $V_{eff}(r)$, is
\begin{eqnarray}
V_{eff}(r)&=&-\left(\frac{\psi_0}{\psi}\right)^{2}\frac{G\zeta\psi
_0}{r}+\left(\frac{\psi_0}{\psi}\right)^{4}\left[\frac{p_{\phi}^2}{2r^2}-\frac{G\zeta\psi
_0 p_{\phi}^2}{c^{2}r^3}\right] \nonumber \\
&-&
\frac{1}{2}\left(\frac{\psi_0}{\psi}\right)^{2}\left[\left(U^{\psi}\right)^2\left(\frac{2G\zeta\psi_0}{c^2
r}-\frac{r^2}{\psi_0^2}\right)-
\left(\frac{r}{\psi_0}\right)^{2}\right]. \label{b7}
\end{eqnarray}

However, we are interested in the study of this potential for
massive test particles on static foliations $\psi=\psi_0=c/H_0$,
such that the dynamics evolves on an effective 4D manifold
$\Sigma_0$. From the point of view of a relativistic observer,
this implies that $U^{\psi}=0$.

\subsection{Geodesics equations for 5D canonical metrics}

We consider a 5D line element $dS^2=g_{ab}(x^c) dx^a dx^b$. We are
interested in studying the geodesics equations on a 5D canonical
metric
\begin{equation}
dS^2= \left(\frac{\psi}{\psi_0}\right)^2 ds^2 - d\psi^2,
\end{equation}
where $ds^2=h_{\alpha\beta}(x^{\mu}) dx^{\alpha} dx^{\beta}$, such
that in the absence of external forces the 5D geodesic equation is
\begin{equation}\label{geo}
\frac{d^2 x^a}{dS^2} + \Gamma^a_{bc} \frac{d x^b}{dS} \frac{d
x^c}{dS} =0.
\end{equation}
For a test particle in a time-like geodesic we must require
\begin{equation}\label{aaa}
g_{ab} U^a U^b =1,
\end{equation}
such that the velocity components are $U^c={dx^c \over dS}$\footnote{The case of 5D null geodesics have been studied in earlier
works\cite{se}.}. To study the effective 4D geodesic equations on a
hypersurface obtained after making a constant foliation
$\psi=\psi_0$, we shall decompose (\ref{geo}) in the geodesic
equations
\begin{eqnarray}
\frac{d^2 x^{\mu}}{ds^2} & + & \bar{\Gamma}^{\mu}_{\alpha\beta}
\frac{d x^{\alpha}}{ds} \frac{d x^{\beta}}{ds} = -
\frac{d^2s}{dS^2} \left(\frac{ds}{dS}\right)^{-2}
\frac{dx^{\mu}}{ds} - 2 \frac{1}{\psi_0} \delta^{\mu}_{\nu}
\frac{dx^{\nu}}{ds}
\frac{d\psi}{ds}, \label{aa} \\
\frac{d^2 \psi}{ds^2} & + & {\Gamma}^{4}_{\alpha\beta} \frac{d
x^{\alpha}}{ds} \frac{d x^{\beta}}{ds}=- \frac{d^2s}{dS^2}
\left(\frac{ds}{dS}\right)^{-2} \frac{d\psi}{ds}, \label{bb}
\end{eqnarray}
where
\begin{equation}\label{11}
\frac{ds}{dS} =  \left[\left( \frac{\psi}{\psi_0}\right)^2 -
\left(\frac{d\psi}{ds}\right)^2 \right]^{-1/2}.
\end{equation}
Deriving the last expression with respect to $S$, we obtain
\begin{equation}\label{bbb}
\frac{d^2s}{dS^2} \left(\frac{ds}{dS}\right)^{-2} = -
\left(\frac{ds}{dS}\right)^2 \frac{d\psi}{ds}
\left[\frac{\psi}{\psi_0^2} - \frac{d^2\psi}{ds^2}\right].
\end{equation}
Using (\ref{aaa}) and (\ref{bbb}) in (\ref{aa}) and (\ref{bb}), we
obtain
\begin{eqnarray}
\frac{d^2 x^{\mu}}{ds^2} & + & {\Gamma}^{\mu}_{\alpha\beta}
\frac{d x^{\alpha}}{ds} \frac{d x^{\beta}}{ds} =
\frac{dx^{\mu}}{ds}\frac{d\psi}{ds} \left(\frac{ds}{dS}\right)^2
\left[\frac{\psi}{\psi_0^2} - \frac{d^2\psi}{ds^2}\right], \label{au}\\
\frac{d^2 \psi}{ds^2} & + &  \frac{\psi}{\psi_0^2} =
\left(\frac{ds}{dS}\right)^2
\frac{d\psi}{ds}\left[\frac{\psi}{\psi_0^2} -
\frac{d^2\psi}{ds^2}\right] . \label{bu}
\end{eqnarray}
Using (\ref{11}) and (\ref{bbb}) in (\ref{au}) we obtain that
the right-hand side of (\ref{au}) becomes null, so that the system
(\ref{au})-(\ref{bu}) finally becomes
\begin{eqnarray}
\frac{d^2 x^{\mu}}{ds^2} & + & {\Gamma}^{\mu}_{\alpha\beta}
\frac{d x^{\alpha}}{ds} \frac{d x^{\beta}}{ds} =0, \label{aau}\\
\frac{d^2 \psi}{ds^2} & + &  \frac{\psi}{\psi_0^2} =
\frac{2}{\psi} \left(\frac{d\psi}{ds}\right)^2 . \label{bbu}
\end{eqnarray}
The solution of this set of equations is
\begin{equation}\label{solu}
\psi(s) = - \frac{2 e^{-s/\psi_0}}{\psi_0
\left[C_1\,e^{-2s/\psi_0} + C_2\right]}.
\end{equation}
We are interested in studying the induced dynamics of observers who
moves on the hypersurface $\Sigma_0$, resulting from setting a constant
foliation $\psi(s)=\psi_0$. In the next section we shall consider
this case which will be relevant to the study of planetary
dynamics on an effective 4D Schwarzschild-de Sitter space-time.

\section{Physics on the 4D manifold $\Sigma_0$ in the solar system}

Now we consider the static foliation $\lbrace\Sigma _{0}
:\psi=\psi _{0}\rbrace$ on (\ref{a1}). In this case we obtain the
effective 4D line element
\begin{equation}\label{a2}
dS^{2}_{ind}=c^{2}f(r)dt^{2}-\frac{dr^{2}}{f(r)}-r^{2}\,\left[d\theta
^{2}+sin^{2}(\theta) \,d\phi ^{2}\right],
\end{equation}
which is known as the Schwarzschild-de Sitter metric. From the
relativistic point of view, observers that are on $\Sigma_0$ move
with $U^{\psi}=0$. We assume that the induced matter on $\Sigma_0$
can be globally described by a 4D energy momentum tensor of a
perfect fluid $T_{\alpha\beta}=(\rho c^2
+P)U_{\alpha}U_{\beta}-Pg_{\alpha\beta}$, where $\rho(t,r)$ and
$P(t,r)$ are respectively the energy density and pressure of the
induced matter, such that
\begin{equation}\label{a7}
P= -\rho c^2 =-\frac{3c^4}{8\pi G}\frac{1}{\psi _{0}^2},
\end{equation}
which corresponds to a vacuum equation of state. The
energy density of induced matter is denoted $\rho$. Because we are interested
in studying the orbits of some planets of our solar system, we shall
consider that the main gravitational source is the Solar mass
$M_{\odot}\equiv \zeta\psi_0$ and radius $r_0$. We shall assume
that we live on the 4D hypersurface $\Sigma _{H_0}:\psi
_{0}=cH_{0}^{-1}$, $H_0$ and $G\zeta \le 1/ (2 \sqrt{27})\simeq
0.096225$, being $H_{0}=73\,\frac{km}{sec}Mpc^{-1}$ the present
day Hubble constant.

When one takes $U^{\psi}=0$, the induced potential $V_{ind}(r)$ on
the hypersurface $\Sigma _0$ is given by
\begin{equation}\label{b9}
V_{ind}(r)=-\frac{G
M_{\odot}}{r}+\frac{p_{\phi}^{2}}{2r^2}-\frac{G
M_{\odot}}{c^2}\frac{
p_{\phi}^2}{r^3}-\frac{1}{2}\left(\frac{r}{\psi_0}\right)^{2}.
\end{equation}
The first two terms on the right hand side of (\ref{b9})
correspond to the classical potential, the third term is the usual
relativistic contribution and the last term is a new contribution
coming from 5D metric solution (\ref{a1}). The acceleration
associated with the induced potential (\ref{b9}) reads
\begin{equation}\label{in2}
a=-\frac{GM_{\odot}}{r^2}+\frac{p_{\phi}^2}{r^3}-\frac{3GM_{\odot}}{c^2}\frac{p_{\phi}^2}{r^4}+\frac{r}{\psi
_0^2}.
\end{equation}
By expressing (\ref{b6}) as a function of the angular
coordinate, $\phi$ (indeed assuming $1/u=r=r(\phi)$), we obtain
\begin{equation}\label{b12}
\left(\frac{du}{d\phi}\right)^{2}+(1-\frac{2G M_{\odot}}{c^2}
u)(p_{\phi}^{-2}+u^{2})-p_{\phi}^{-2}(u\psi_0)^{-2}=c^{-2}p_{t}^{2}p^{-2}_{\phi}+\psi_{0}^{-2}.
\end{equation}
This equation of the orbit is almost the same that the one usually
obtained in the 4D general theory of relativity for a
Schwarzschild-de Sitter metric. However, notice that here the
cosmological constant is well determined by the constant
$\psi^{-2}_0=H^2_0/c^2$, and not any constant of arbitrary
signature (as in 4D general relativity). In other words, in our
formalism the cosmological constant is determined geometrically by
the foliation.

\subsection{Effective geodesics equations on the 4D hypersurface}

If we require that $S(s)=s$, we must place (\ref{solu}) in (\ref{11}). Hence, after taking 
a constant
foliation $\psi=\psi_0$, the solution for $S(s)$ is
\begin{equation}\label{21}
S(s)=s=-\frac{\psi_0}{2} {\rm ln}\left(-\frac{C_2}{C_1}\right).
\end{equation}
In this case both (\ref{aau}) and (\ref{bbu}) evaluated
on the foliation $\psi=\psi_0$ are free of sources
\begin{eqnarray}
\frac{d^2 x^{\mu}}{ds^2} & + & \bar{\Gamma}^{\mu}_{\alpha\beta}
\frac{d x^{\alpha}}{ds} \frac{d x^{\beta}}{ds} =0, \label{aaau}\\
\frac{d^2 \psi}{ds^2} & + &  \frac{1}{\psi_0} = 0, \label{bbbu}
\end{eqnarray}
where $\bar{\Gamma}^{\mu}_{\alpha\beta} =
\left.{\Gamma}^{\mu}_{\alpha\beta}\right|_{\psi=\psi_0}$. Finally,
we must additionally require that $C_2=-1/(\psi_0^4\,C_1)$ in
order to obtain $\psi(s)=\psi_0$ in (\ref{solu}). Notice that
there are no extra force components in (\ref{aaau})
and (\ref{bbbu}).

\section{Calculation of the perihelion advance}

In order to study the advances of perihelions for massive test
particles in the solar system we consider (\ref{b12}). After making $u(\phi)= 4/r_s\,v(\phi)$ and
$M={r_{s}}/{256p^{2}_{\phi}\psi_{0}^{2}}$, we obtain
\begin{equation}\label{a18}
{v^2}\left(\frac{dv}{d\phi}\right)^{2}=4v^5-v^4+\frac{v^{3}r_{s}^2}{4p_{\phi}^{2}}-v^2[\frac{r_{s}^2}{16p_{\phi}^{2}}
-\nonumber\frac{r_{s}^2}{16}{[p_{t}^{2}p^{-2}_{\phi}+\psi_{0}^{-2}]}]+M,
\end{equation}
where 
\begin{equation}
P_5(v)=4v^5-v^4+\frac{v^{3}r_{s}^2}{4p_{\phi}^{2}}-v^2[\frac{r_{s}^2}{16p_{\phi}^{2}}
-\nonumber\frac{r_{s}^2}{16}{[p_{t}^{2}p^{-2}_{\phi}+\psi_{0}^{-2}]}]+M.
\end{equation}
The half-period of the orbit will be
\begin{equation}\label{a19}
\phi +\phi_0=\int_{e_{1}}^{e_{2}}\frac{v}{\sqrt{P_{5}(v)}}\,dv,
\end{equation}
where $e_{1}$ and $e_{2}$ are the real and positive roots of
$P_{5}(v)$.

The advance of the perihelion for the orbits will be given by two
times the difference between $\pi$ and the angle described by the
orbit in (\ref{a19})
\begin{equation}\label{a21}
\Delta_{\phi}^{MI}=2\pi-2\int_{e_{1}}^{e_{2}}\frac{v}{\sqrt{P_{5}(v)}}\,dv.
\end{equation}
It must be noted that $M\ll 1$. In order to calculate the
integral in (\ref{a21}), we shall make the following expansion of
${v}/{\sqrt{P_{5}(v)}}$
\begin{equation}\label{a22}
\frac{v}{\sqrt{P_{5}(v)}}|_{M\ll
1}\simeq=\frac{1}{\sqrt{P_{3}(v)}}-\frac{M}{2\sqrt{\left[P_{3}(v)\right]^3}}+\frac{3M^3}{8\sqrt{\left[P_{3}(v)\right]^5}}+...\,
,
\end{equation}
with $P_5(v)=v^2 P_3(v)+M$, and
\begin{equation}\label{a24}
P_{3}(v)=4v^3-v^2+\frac{r_{s}^2v}{4p_{\phi}^{2}}-\frac{r_{s}^2}{16p_{\phi}^{2}}-\frac{r_{s}^2}{16}{[p_{t}^{2}p^{-2}_{\phi}+\psi_{0}^{-2}]}.
\end{equation}
Notice that all the terms in the series are integrable. Finally, if
we make the substitution $v(\phi)=w(\phi) + 1/12$, we obtain the
result
\begin{equation}\label{a30}
\Delta_{\phi}^{MI}=\phi_1 +\phi_2= \int_{\epsilon_1}^{\infty}
\frac{dw}{\sqrt{4 w^3 - g_2 w - g_3}} - \frac{M}{2}
\int_{\epsilon_1}^{\infty}
\frac{dw}{\left(w+\frac{1}{2}\right)^2\sqrt{4 w^3 - g_2 w - g_3}}+
...\, ,
\end{equation}
with
\begin{eqnarray}
\phi_1& = & \int_{\epsilon_1}^{\infty} \frac{dw}{\sqrt{4 w^3 -
g_2 w - g_3}}, \label{a331} \\
\phi_2 & = & - \frac{M}{2} \int_{\epsilon_1}^{\infty}
\frac{dw}{\left(w+\frac{1}{2}\right)^2\sqrt{4 w^3 - g_2 w - g_3}},
\label{a332}
\end{eqnarray}
where $g_2$ and $g_3$ are the invariants of Weierstrass
\begin{eqnarray}
g_2 &= & \frac{1}{12} - \frac{r^2_s}{4 p^2_{\phi}}, \\
g_3 & = & \frac{1}{216} + \frac{r^2_s}{16} \left[
\frac{1}{p^2_{\phi}} \left( 1 - \frac{p^2_t}{c^2} \right) +
\frac{1}{\psi^2_0} \right] - \frac{r^2_s}{48 \, p^2_{\phi}},
\end{eqnarray}
and $e_1=\epsilon_1+1/12$, such that $P_3(w=\epsilon_1)=0$. The
constants $p_t$ and $p_{\phi}$ are the two free parameters of the
theory and they are related to the energy by mass
unit, $E=c\, p_t$, and the angular moment by mass unit, $L_M=c\,
p_{\phi}$, such that the invariants of Weierstrass hold
\begin{eqnarray}
g_{2}& = & \frac{1}{12}-\frac{r_{s}^2c^2}{4 L_{M}^{2}},
\label{a31} \\
g_{3} &=
&-\frac{r_{s}^2c^2}{48L_{M}^{2}}+\frac{1}{216}+\frac{r_{s}^2c^2}{16L_{M}^{2}}-\frac{r_{s}^2}{16}{\left[\frac{E^{2}}{c^2L^{2}_{M}}+\psi_{0}^{-2}\right]}.
\label{a32}
\end{eqnarray}
Furthermore, because $0 < r < \infty$, the range of validity of
$w(\phi)$ is: $-1/12 < w < \infty$.

\subsection{Limit case with $\psi_{0}\rightarrow\infty$}

Because $\psi_0=c/H$, the case with zero cosmological constant
corresponds to the limit case $\psi_0 \rightarrow \infty$. Notice
that $H$ is the Hubble parameter which is experimentally
determined so that the foliation $\psi=\psi_0$ is given physical
parameters. If we take this limit in (\ref{a31}) and (\ref{a32})
we obtain exactly the same solution as (\ref{a30}), but the
invariant of Weierstrass $\hat{g_{2}}$ and $\hat{g_{3}}$
\begin{eqnarray}
\hat{g_{2}}&=&\frac{1}{12}-\frac{r_{s}^2c^2}{4 L_{M}^{2}}, \\
\hat{g_{3}}&=&-\frac{r_{s}^2c^2}{48L_{M}^{2}}+\frac{1}{216}+\frac{r_{s}^2c^2}{16L_{M}^{2}}-\frac{r_{s}^2}{16}{\left[\frac{E^{2}}{c^2L^{2}_{M}}\right]}.
\end{eqnarray}
These expressions are in agreement with the results obtained when
we use the standard 4D formalism for general relativity.

\section{Numerical results}

With the aim to illustrate the formalism we shall calculate the
advance for the perihelions of Mercury, the Earth and Pluto. We
shall use for our calculations the respective values for the
Schwarzschild radius ($r_s$), the speed of light ($c$) and the
Hubble radius ($c/H$): $r_{s}=2.95325008\times10^5$ cm,
$c=2.9979245800\times 10^{10}$ cm/seg and $c/H=1.2701000000\times
10^{28}$ cm. In all cases we shall consider that the angular
moment by mass unit is given by $L_M = v_p r_p$, such that $v_p$
and $r_p$ are the velocity and distance, respectively, of the planet
at the perihelion.

\subsection{Mercury}

The orbital period of Mercury is
$87.9695$ Earth days. Its angular moment by mass unit is
$L_{M}=2.71308044481\times 10^{19}$ $cm^2/seg$ and the energy by
mass unit being given by $E=2.99792454178\times 10^{10}$ $cm/seg$.
The only finite real root on the physical domain is
$\epsilon_1=0.166666640044$. Using (\ref{a19}), we can
calculate the half-period: $\phi=3.14159290450$. It is very
important to notice that the result of the second integral in
(\ref{a332}) is negligible: $\phi_2=-9.71527962041 \times
10^{-52}$, so that the advance of the perihelion results given
totally by the first integral (\ref{a331}):
$\Delta_{\phi}^{MI}=42.9773350296$ arcseg/century. This value is
in very good agreement with observations:
$\left.\Delta_{\phi}^{MI}\right|_{exp}=42.98 \pm 0.04$ and with
predictions of general relativity\cite{Kra}.

\subsection{Earth}

The Earth is densest and
fifth-largest of the eight planets in the Solar System. Its
angular moment per mass units is $L_{M}=4.52332500000\times
10^{19}$ $cm^2/seg$ and and its energy per mass units is
$E=2.99792457200\times 10^{10}$ $cm/seg$. The only finite real
root on the physical domain is $\epsilon_{1}= 0.166666657089$.
Therefore, for an orbital period of $365$ days, the half-period can
be calculated from eq. (\ref{a19}): $\phi=3.14159274386$. The
advance of the perihelion of the Earth can be calculated from the
first integral (\ref{a331}): $\Delta_{\phi}^{MI}=3.72390481198$
arcseg/century. This value agree with the experimentally observed
value. Because in the case of Mercury the second integral
(\ref{a332}) is very small: $\phi_2=-3.49514238656\times
10^{-52}$.

\subsection{Pluto}

It is well known that Pluto it is not a true planet. It is the
second most massive known dwarf planet, after Eris. In this case
the angular moment by mass unit is
$L_{M}=2.7025100000\times10^{19}$ $cm^2/seg$ and the energy by
mass unit that we use is $E=2.9979245418\times10^{10}$ $cm/seg$,
so that the root in the physical domain takes the value
$\epsilon_{1} = 0.16666666639$. Pluto has an orbital period of
$247.08$ terrestrial years so that the half-period is
$\phi=3.1415926561$. This value being given by the first integral
(\ref{a331}), because the second one (\ref{a332}) is two orders of
magnitude smaller than the other two cases:
$\phi_2=-9.791421275\times 10^{-54}$. With these values we can
calculate the advance of the perihelion, which takes the value
$\Delta_{\phi}^{MI}=0.000417$ arcseg/century.

\section{Final comments}

Induced matter theory\cite{We92, Over97, Liko04, PdL01, Se02}, has been
of much interest in recent years and the exploration of the
geodesic equations from a 5D vacuum is an important topic of this
theory\cite{Se01}. In this paper we have re-examined
this topic to apply to possible
applications of ETGR to orbits like planetary orbits in our
solar system. ETGR has been proposed some years ago\cite{MB} and has been studied in the framework
of astrophysical\cite{Pitu} and cosmological\cite{JCAP10, PL14}
applications. However, the possible applications are
not limited to these kinds of orbits. A very important result is
the particular solution with $S(s)=s$ described in Sect. IIIa, for
which there are no extra force components due to the foliations on
the extra dimension [see (\ref{aaau}) and (\ref{bbbu})].

We have studied analytically the advances for the perihelions for
Mercury, the Earth and Pluto. This work was the first to use ETGR, where the
cosmological constant is determined by the foliation
$\psi=\psi_0=c/H$, so once the Hubble constant, 
is experimentally determined, we have the cosmological constant:
$\Lambda= 3/\psi^2_0=3H^2$. In our calculations we have not
considered the quadrupolar moment of the Sun, which may be
important for the orbit of Mercury\cite{Pi}.

This method can be used to calculate
other orbits of comets with large period that come from the
Oort cloud. Some of these comets, as for example, the Ison comet,
pass very close to the Sun and therefore are subject to an intense
gravitational field\cite{Ma}.

\section*{Acknowledgements}

\noindent S. Ridao and R. Avalos and M. D. De Cicco acknowledge
UNMdP for financial support. M. Bellini acknowledges CONICET
(Argentina) and UNMdP for financial support.

\end{document}